\RequirePackage{fix-cm}
\documentclass[smallextended]{svjour3}       % onecolumn (second format)
\smartqed  % flush right qed marks, e.g. at end of proof
\usepackage{graphicx}
\usepackage{amsmath,amssymb}
\begin{document}

\title{Triple path to the exponential metric}
% \subtitle{Do you have a subtitle?\\ If so, write it here}


\author{Maxim Makukov        \and
  Eduard Mychelkin
}


\institute{Maxim Makukov \at
  Fesenkov Astrophysical Institute \\
  050020, Almaty, Republic of Kazakhstan \\
              \email{makukov@aphi.kz}           %  \\
           \and
Eduard Mychelkin \at
Fesenkov Astrophysical Institute \\
  050020, Almaty, Republic of Kazakhstan \\
              \email{mychelkin@aphi.kz}           %  \\
}

\date{Received: date / Accepted: date}

\maketitle

\begin{abstract}
  The exponential Papapetrou metric induced by scalar field conforms to
  observational data not worse than the vacuum Schwarzschild solution. Here, we
  analyze the origin of this metric as a peculiar space-time within a wide class of
  scalar and antiscalar solutions of the Einstein equations parameterized by
  scalar charge. Generalizing the three families of static solutions obtained by
  Fisher (1948), Janis, Newman \& Winicour (1968), and Xanthopoulos \& Zannias
  (1989), we prove that all three reduce to the same exponential metric provided that scalar charge is equal to central mass,
  thereby suggesting the universal character of such background scalar field.
\end{abstract}

\section{Introduction}

Understanding the universe evolution depends on the correct answer to the
question -- what kind of physical vacuum do we live in? In general relativity
there is a mainstream in the study of various effects and processes related to
known vacuum solutions of the Einstein equations, such as black holes and
gravitational waves. On the other hand, similar phenomena may arise due to (or
be influenced by) some background scalar field $\phi \left( x^\alpha  \right)$
suggested, in particular, by the dark side of the universe. E.g., the presence
of such background scalar field affects whether one will observe regular black
holes or, rather, compact objects without horizons \cite{Janis1968,agnese,mm18}.

 In this regard, the exponential spherically symmetric Papapetrou metric
 \cite{papa}:
\begin{equation}
  ds^2 =  e^{-2\phi(r)} {dt}^2 - e^{2 \phi(r)} \left( {dr}^2 + r^2 d\Omega^2
  \right),
  \label{first}
\end{equation}
with $\phi(r) = M/r$, represents a viable alternative to pure vacuum
approach \cite{mm18} (we adopt $\phi > 0$). The fact that the source for the
exponential metric represents a minimal scalar field in
antiscalar regime (see below) was first pointed out by Yilmaz
\cite{yilmaz}. Note also that the exponential metric satisfies the ``Papapetrou
ansatz'' 
${g}_{\mu\nu} = {g}_{\mu\nu}\left( \phi \left( x^\alpha  \right)  \right)$,
which picks out the class of metrics whose coefficients depend on coordinates
solely through the background scalar field. As evident from (\ref{first}), this
metric is horizon-free.

In the isotropic form (\ref{first}) of the metric, both finite
and infinitesimal spatial intervals have the same scale factor. Besides, this
spatially-conformal factor is reciprocal to that of the time interval, so that in the weak field approximation this automatically leads to the Newtonian gauge widely used, in particular, in cosmological applications. The potential in exponents in (\ref{first}) plays a key
  role in that it allows to make quantitative estimations of effects generated by
  scalar background. Noteworthy, the exponential form of metric coefficients
  might be justified by fundamental physical considerations related, in
  particular, to the propagation of light in static fields \cite{Rindler}.
 
Another peculiar feature of the exponential metric is that it leads to the
Newtonian potential as exact solution of the corresponding Klein-Gordon
equation, $g^{\mu\nu}\phi_{;\mu\nu}=0$, so this potential might be
used not only
in weak-field approximations, but in arbitrarily strong fields \cite{mm19}. Besides, with the exponential metric one may build the general approach to thermodynamics of scalar background, which  generates the well-known black-hole thermodynamic relations at the gravitational radius scale \cite{mm18}. 

 In spite of all these features, the credibility of Papapetrou's solution might
 be limited as the origin and nature of scalar field appears problematic. The
 problem of scalar field diagnostics and determination of scalar charge value
 was posed in \cite{Virbhadra1998}. Acting in this direction, we generalize
 three families of scalar solutions found by Fisher (1948) \cite{fisher}, Janis,
 Newman \& Winicour (1968) \cite{Janis1968} and Xanthopoulos \& Zannias (1989)
 \cite{xz} from scalar to antiscalar regime, and show that the
 exponential  Papapetrou metric follows in each case uniquely through the identification (in
 corresponding units) of scalar charge with the central mass.

\section{The three static metrics with scalar field}%
\label{sec:unif_}
\subsection{Scalar and antiscalar regimes}
We approach the Einstein-Klein-Gordon system with minimal spherically-symmetric static scalar field in a generic way, which includes both scalar and antiscalar regimes simultaneously. To this end, in all solutions we employ the same generalized representation of the master $\gamma$-factor parameterized by the scalar charge. 

Antiscalarity implies the opposite sign of the scalar field energy-momentum tensor (EMT) in the field equations. In terms of Lagrangian this reads
\begin{equation}
	L = \frac{1}{16\pi} \left( R - 2\epsilon \phi_\alpha \phi^\alpha\right) \qquad \Rightarrow \qquad
	{G}_{\mu\nu} = 8 \pi \epsilon  {T}^{SF}_{\mu\nu},
	\label{EE}
\end{equation}
where the scalar EMT, ${T}_{\mu\nu}^{SF} = \frac{1}{4 \pi} \left( \phi_\mu \phi_\nu - \frac{1}{2}{g}_{\mu\nu} \phi_\alpha \phi^\alpha \right)$, is quadratic with respect to $\phi_\alpha = \nabla_\alpha \phi = \phi_{,\alpha}$, and the sign-factor is defined as $\epsilon = \{ +1, 0, -1\}$ for scalar, vacuum and antiscalar regimes, correspondingly. 

It stands to reason that the ordinary matter and radiation cannot be used with
negative sign of the energy-momentum tensor in order to avoid problems with
energy dominance and stability. As for media with an exotic equation of
state, such as cosmological $\Lambda$-term or minimal scalar
field\footnote{The cosmological term (both in de Sitter and anti-de Sitter
    regimes) is characterized by the equation of state
    $p= -\varepsilon$, while minimal scalar field (both in standard and
    antiscalar regimes) corresponds to $p=\varepsilon$
  for time-like gradient, and to $p =
-\varepsilon/3$ for space-like gradient.}, the choice of the sign for EMT,
in general, should be defined according to observational data following (or
expected from) the problem under consideration.

In contrast to the case of de Sitter and anti-de Sitter
$\Lambda$-vacuum solutions of the Einstein equations, which are emphatically
distinct (embedded into higher-dimensional space, dS and AdS metrics
lead to different topologies), the solutions of (\ref{EE}) in
scalar and antiscalar regimes functionally coincide after absorbing the
sign-factor $\epsilon$ into the master $\gamma$-factor which depends on scalar charge, as described below (see also \cite{bronnikov} and \cite{mm18}).

\subsection{Fisher (1948)}
Fisher \cite{fisher} was the first to consider the static solution of the field equations with scalar field in the context of the search for new metrics to describe the behavior of meson fields in gravity. In massless approximation the solution of the Einstein equations for minimal scalar field ($\epsilon = 1$),
\begin{equation}
	{G}_{\mu\nu} = \frac{8 \pi G}{c^4} \,{T}^{SF}_{\mu\nu},
		\label{ein}
\end{equation}
and of the corresponding Klein-Gordon equation,
\begin{equation}
 {{{T}_{\mu}}^{\alpha}}_{;\alpha} = 0 \quad \Rightarrow \quad 	\square \phi = \frac{1}{\sqrt{-g}}\frac{\partial \left( \sqrt{-g} g^{\mu \alpha}\partial_\alpha \phi\right)}{\partial x^\mu} = 0,
 	\label{kg}
\end{equation}
for static spherical symmetry was obtained by Fisher in
curvature coordinates:
\begin{equation}
ds^2 = e^{\nu}c^2 {dt}^2 - e^{\lambda} {dR}^2 - R^2 d\Omega^2.
\label{CurvCoord}
\end{equation}
With that, Fisher introduced a specific variable 
\begin{equation}
Z=R e^\frac{\nu - \lambda}{2},
	\label{zet}
\end{equation}
and the problem was reduced to the determination of the explicit dependence $Z(R)$.
The expression (\ref{zet}) was deduced from the first integral of the
Klein-Gordon equation (\ref{kg}), 
\begin{equation}
  \phi' = -\frac{const}{R^2}  e^{\frac{\lambda - \nu}{2}},
  \label{kgfisher}
\end{equation}
and the symmetry condition in curvature coordinates:
\begin{equation}
  (R^2  e^{\nu-\lambda})' = 2R e^\nu.
\end{equation}
Ultimately, the solution of the Einstein equations might be represented in the following form: 
\begin{eqnarray}
e^\nu &=& \frac{Z^2}{R^2 } 	\left( Z - Z_0 \right) 	\left( Z + Z_1 \right) = \left( \frac{Z-Z_0}{Z+Z_1} \right)^\gamma,\nonumber\\
e^\lambda &=&  \frac{ R^2}{Z^2} \left(\frac{Z-Z_0}{Z+Z_1}\right)^\gamma = \left( Z - Z_0 \right) 	\left( Z + Z_1 \right) .
\label{FisherMetric}
\end{eqnarray}
Taking (\ref{kgfisher}) into account, the associated scalar field  is found to be
\begin{equation}
\tilde{\phi}(R) = \frac{c^2\sigma}{2\sqrt{G^2 M^2 + a^2 c^2}}\ln \frac{Z + Z_1}{Z - Z_0},
\label{FishPot}
\end{equation}
with $a^2 =G \sigma^2/c^2$.  Here all values prove to be defined only up to the
unknown quantity $Z$ satisfying the condition
\begin{equation}
R(Z)^2 = 	\left( Z - Z_0 \right)^{1-\gamma} 	\left( Z + Z_1 \right)^{1+\gamma} ,
\label{Zcond}
\end{equation}
with 
\begin{eqnarray}
Z_0 &=& c^{-2} \left( \sqrt{G^2 M^2 + c^2 a^2 } - G M \right), \quad Z_1 = c^{-2} \left( \sqrt{G^2 M^2 + c^2 a^2 } + G M \right),\nonumber\\
\gamma &=& \frac{G M}{\sqrt{G^2 M^2 + c^2 a^2 }} = \frac{M}{\sqrt{M^2 + \sigma^2/G}},  
\label{z0z1}
\end{eqnarray}
where $\sigma$ is an arbitrary constant which has the meaning of scalar charge
with the dimensionality of $\sqrt{ G } M$. Having in mind the crucial role of
$\gamma$-factor (due to its reversible dependence on scalar charge,
$\gamma=\gamma(\sigma)$) it is convenient to write the dimensionless counterpart
of potential (\ref{FishPot}) with the aid of (\ref{z0z1}) as:
\begin{equation}
\phi(R) =\tilde{\phi}(R)\frac{\sqrt{G}}{c^2} = \frac{1}{2} \sqrt{1-\gamma^2}\ln\frac{Z + Z_1}{Z - Z_0}
\label{FisherPot}
\end{equation}
(in units $G=c=1$ expressions (\ref{FishPot}) and (\ref{FisherPot}) coincide). 

Provided that (\ref{Zcond}) cannot be reversed to $Z(R)$ analytically for
arbitrary $\gamma$ (apart from certain limiting cases), the Fisher solution (in
contrast to the JNW- and XZ-solutions, the latter to be discussed below) proves to be not closed and is defined only up to the dependence $R(Z)$. So, it is not suitable for direct general relativistic computations (e.g. finding the Ricci and Kretschmann invariants). As will be shown, this appears as an artifact related to curvature coordinates.

\subsection{Janis-Newman-Winicour (1968)}
In 1968 Janis, Newman and Winicour envisaged the same problem in the context of
spacetime singularities, an issue relatively fresh at that time. With that, they
construct the coordinates where, under limiting procedure, the exterior
Schwarzschild event horizon should map onto a point rather than a sphere. The
resulting metric as the solution of (\ref{ein}) (hereafter we use the units $G=c=1$) might be represented in the form \cite{Janis1968} (see also \cite{Virbhadra1997}):
\begin{equation}
ds^2 = \left(1-\frac{2M}{\gamma  r}\right)^{\gamma }{dt}^2 -\left(1-\frac{2M}{\gamma  r}\right)^{-\gamma }{dr}^2 - \left(1-\frac{2M}{\gamma  r}\right)^{1-\gamma } r^2 d\Omega^2.
\label{JNW}
\end{equation}
This metric is singular at $r=\gamma/(2M)$. However, as might be verified, it does
not correspond to a horizon, since the surface area of a sphere 
at this radius proves to be zero in this metric (see, e.g., \cite{agnese}).

As was already noted, a peculiar feature of the Einstein equations (\ref{EE}) is that solutions in both scalar and antiscalar regimes in fact coincide up to the sign-factor $\epsilon$ entering now into
 $\gamma$-parameter which specifies the family of solutions (cf. \cite{bronnikov,mm18}):
\begin{equation} \gamma(\epsilon, \sigma) = M/\sqrt{M^2+ \epsilon \sigma^2},
\label{gamma}
\end{equation}
with $\sigma$ being the scalar charge. It turns out that of all this family of solutions only two values of $\gamma$ might be of special physical interest. These are the Schwarzschild solution in curvature coordinates for $\gamma = 1$,
\begin{equation}
ds^2 =  (1-2M/r)dt^2 - (1-2M/r)^{-1} {dr}^2 -  r^2 d\Omega^2,
\label{Schwarz}
\end{equation}
and the antiscalar Papapetrou solution \cite{papa} in isotropic coordinates for $\gamma~\to~\infty$ (note that this limit automatically implies antiscalarity),
\begin{equation}
ds^2 =  e^{-2M/r} {dt}^2 - e^{2M/r} \left( {dr}^2 + r^2 d\Omega^2 \right).
\label{Papa0}
\end{equation}
For appropriate comparison of these two metrics (and corresponding effects) one should recast the Schwarzschild metric into isotropic coordinates using the transformation $r \mapsto r\left(1+{M}/(2r)\right)^2$. Evidently, the case $\gamma = 1$ corresponds to classical vacuum with $\epsilon = 0$ or/and a tantamount condition $\sigma = 0$, while the limit $\gamma\to\infty$ corresponds to antiscalar case $\epsilon=-1$ with the value of $\sigma$ uniquely fixed at $\sigma=M$. The latter, in its turn, leads to the conclusion that (at least for the given family of solutions) \emph{exactly masses serve as scalar field sources}. It is reasonable to ask if this feature is an artefact of a certain choice of family under consideration. We will address this question below.

The Lagrangian in (\ref{EE}) leads also to the Klein-Gordon equation $\Box \phi
= 0$ with the static solution in the general metric (\ref{JNW}),
\begin{equation}
  \phi =-\frac{1}{2} \sqrt{\epsilon\left(1- \gamma^2\right)}\ln \left( 1 - \frac{2M}{\gamma r}  \right),
\label{KGsol}
\end{equation}
which in antiscalar case for $\gamma \to \infty$ reduces exactly to the Newtonian scalar field,
\begin{equation}
\phi \mapsto \lim_{\gamma \to \infty} \phi \big\rvert_{\epsilon=-1}   = \frac{M}{r},
\label{NewtPot}
\end{equation}
as it should be for the exponential metric. Besides, one can see that in scalar case factor $\gamma$ might change from 0 to 1, while in antiscalar regime it varies from 1 to $\infty$, so the complete interval is $[0, \infty)$.

\subsection{Xanthopoulos \& Zannias (1989)}
Xanthopoulos \& Zannias (XZ) \cite{xz}  had envisaged the scalar Einstein equations within the context of multidimensional approach and adopting isotropic coordinates from the start. Restricted to the 4-dimensional case, 
\begin{equation}
ds^2 = e^{\nu(r)} {dt}^2 - e^{\mu(r)}\left( {dr}^2 + r^2
	d\Omega^2\right),
\label{XZmetric}
\end{equation}
the XZ-solution of the Einstein-Klein-Gordon system is found
to be:
\begin{eqnarray}
e^\nu &=& \left(  \frac{1-\frac{r_0}{r}}{1+\frac{r_0}{r}}\right)^{2\gamma},\nonumber\\
e^\mu &=&  \left(
	\frac{1-\frac{r_0}{r}}{1+\frac{r_0}{r}}\right)^{-2\gamma}\left(
	1-\frac{r_0^2}{r^2}\right)^2,
\label{XZ}
\end{eqnarray}
with the associated scalar field
\begin{equation}
	\phi = \sqrt{\epsilon\left(1-\gamma^2\right)}\ln\frac{1-\frac{r_0}{r}}{1+\frac{r_0}{r}},
\label{XZpot}
\end{equation}
where $r_0 =  M/(2 \gamma)$ and $\gamma$ is the master parameter. Substituting this into the first integral of
the Klein-Gordon equation,
\begin{equation} 
	\dot{\phi} = \frac{\sigma}{r^2}
	e^{-\frac{\nu+\mu}{2}},
\end{equation}
we obtain the same parameter $\gamma(\sigma, \epsilon) = {M}/{\sqrt{M^2 +
\epsilon\sigma^2}}$ as in the JNW case (see (\ref{gamma})). As before, from
(\ref{XZ}) and (\ref{XZpot}), for vacuum case $\gamma = 1$ we get the
Schwarzschild metric, this time in isotropic coordinates. In antiscalar case ($\epsilon = -1$), the special limit $\gamma \to \infty$ leads again to the Papapetrou metric with positive Newtonian scalar potential $\phi =M/r $, again suggesting the idea that exactly masses serve as scalar field sources acting in antiscalar regime.

\section{Interrelation between the three metrics}

\subsection{Transformation between JNW and XZ}
Unlike the situation with Fisher's metric taken both in scalar and antiscalar
regimes, the transformation law between the JNW and XZ coordinates might be
found simply enough. Indeed, equating (\ref{XZpot}) and the corresponding
expression for the JNW potential (\ref{KGsol}), we deduce the
  rule valid for all $\gamma$ to transfer from the JNW- to XZ-coordinates:
\begin{equation}
r \mapsto r\left( 1 + \frac{M}{2\gamma r} \right)^2.
\label{transformCurvIsotrop}
\end{equation}
Evidently, (\ref{transformCurvIsotrop}) differs from the known transformation
from curvature to isotropic form of the Schwarzschild solution only by the
presence of the $\gamma$-factor parameterized in its turn by scalar charge. This
shows that the JNW- and XZ-solutions represent a direct extension of vacuum
curvature and isotropic Schwarzschild coordinates, correspondingly, onto scalar
and/or antiscalar background. As might be seen from (\ref{transformCurvIsotrop}), in antiscalar limit $\gamma \to \infty$ radial JNW and XZ coordinates coincide, and this explains why both families reduce in that limit to the same metric.

\subsection{Transformations to Fisher}%
\label{sub:xzcurv}

To demonstrate the peculiar character  of curvature coordinates (\ref{CurvCoord}) in application to scalar field
let us first transform the exponential Papapetrou metric (\ref{Papa0}), as a
particular case of the JNW- and XZ-solutions. Since angular coordinates do not change, this leads to the direct substitution
\begin{equation}
R = R(r) = r e^{M/r},
\label{PapaTransform}
\end{equation}
and the inversion of this relation yields the transformation law from isotropic $r$- to curvature $R$-coordinate,
\begin{equation}
r= r(R) = -M/W\left(-\frac{M}{R}\right), 
\label{invers}
\end{equation}
where $W(x)$ is the Lambert function. As a result, one obtains the following
representation of the Papapetrou metric (\ref{Papa0}) in curvature coordinates
(see also in \cite{visser}),
\begin{equation}
	ds^2 = e^{2W\left(-\frac{M}{R}\right)}dt^2 - \left( 1 + W\left( -\frac{M}{R}\right)\right)^{-2}dR^2 - R^2 d\Omega^2,
	\label{PapaLambert}
\end{equation}
simultaneously with the replacement of the Newtonian (in isotropic coordinates)
potential (\ref{NewtPot}) into its curvature counterpart:
\begin{equation}
\phi= \phi(R) = -W(-M/R).
\label{phiLambert}
\end{equation}

However, an attempt to apply the same algorithm to transform the general
isotropic XZ-solution (\ref{XZmetric})-(\ref{XZ}) into curvature coordinates fails because it leads to the condition for $R = R(r)$:
\begin{equation}
	R = R(r)= r\left( 1 - \frac{M}{2r\gamma} \right) \left( 1 + \frac{M}{2r\gamma}  \right) \left(  \frac{  1 - \frac{M}{2 r \gamma} }{   1 + \frac{M}{2r \gamma}}  \right)^{-\gamma}, 
	\label{rhoCurvXZ}
\end{equation}
whose inversion $r(R)$ for arbitrary $\gamma$ is analytically highly
problematic. So, in general, the direct transition from XZ (and JNW) to
curvature coordinates (i.e. to Fisher-type solution) also cannot be done. Only
for special values $\gamma=1$ (zero scalar charge) and $\gamma \to \infty$
(scalar charge equals mass in antiscalar regime), as expected, we recover from
(\ref{rhoCurvXZ}) the familiar reversible transformation rules, $R = r \left( 1 + \frac{M}{2r}  \right)^2$ and $R = r e^{M/r}$.

\subsection{Transformations from Fisher}
\label{sub:fromfisher}
As we have shown, the Fisher approach does not admit, in general, the transition
from curvature coordinates to other coordinates, and the converse is also true.
It is convenient to represent the results (\ref{FisherMetric})-(\ref{FisherPot}) in units $G=c=1$ via generalized $\gamma$-factor, i.e. with $Z_0 = M\left( \frac{1}{\gamma} - 1\right)$, $Z_1 = M \left( \frac{1}{\gamma} + 1  \right)$ and $\gamma = {M}/{\sqrt{M^2 + \epsilon\sigma^2}}$, to obtain:
\begin{align}
R &= R(Z) = {Z} \left( 1 - \frac{M}{Z} \frac{1 - \gamma}{\gamma}\right)^{ \frac{\left( 1 - \gamma \right) }{2}}  \left( 1 + \frac{M}{Z} \frac{1+\gamma}{\gamma} \right)^{\frac{ 	\left( 1 + \gamma \right)}{2}}  , \label{RCurvF}	 \\
	e^\nu &= \frac{Z^2}{R^2} \left( 1 - \frac{M}{Z} \frac{1-\gamma }{\gamma }  \right) \left( 1 + \frac{M}{Z} \frac{1+\gamma }{\gamma }  \right) =  \left(  \frac{ 1 - \frac{M}{Z} \frac{1-\gamma }{\gamma}  }{ 1 + \frac{M}{Z} \frac{1+\gamma }{\gamma}   }   \right)^\gamma, \label{EnuF} \\
	e^\lambda & =\frac{R^2}{Z^2} \left(  \frac{ 1 - \frac{M}{Z}\frac{1-\gamma }{\gamma}  }{ 1 + \frac{M}{Z} \frac{1+\gamma }{\gamma}  }   \right)^\gamma= \left( 1 - \frac{M}{Z} \frac{1-\gamma }{\gamma }  \right) \left( 1 + \frac{M}{Z} \frac{1+\gamma }{\gamma }  \right) , \label{ElaF} \\
	&\phi(Z) = \frac{1}{2}\sqrt{\epsilon\left(1-\gamma^2\right)}\ln \frac{ 1 + \frac{M}{Z} \frac{1+\gamma }{\gamma}  }{ 1 - \frac{M}{Z}\frac{1-\gamma }{\gamma}  }  . \label{phiF}
\end{align}
Evidently, for arbitrary $\gamma$ the relation (\ref{RCurvF})  cannot
analytically be inverted into $Z(R)$, similarly to situation with (\ref{rhoCurvXZ}).

Nevertheless, for the situation when scalar charge equals mass, i.e. in the
antiscalar limit  $\gamma \to \infty$, the generalized Fisher expressions in
(\ref{RCurvF})-(\ref{phiF}) reduce to reversible relations:
\begin{eqnarray}
	R(Z) &\to& \left(M+Z\right)e^{\frac{M}{M+Z}} \label{RZ_}\\
	e^\nu &\to& e^{-\frac{2M}{M+Z}} \label{ENU_}\\
	\phi &\to& \frac{M}{M+Z} \label{phi_}.
\end{eqnarray}
As for $e^\lambda$ (\ref{ElaF}), we follow the standard Tolman transition
algorithm from the curvature ($R$) to isotropic ($r$) coordinate. Then,
substituting the exact relations $e^\lambda(Z)$ (\ref{ElaF}) and $R(Z)$ (\ref{RCurvF}) into
\begin{equation}
	\int \frac{dr}{r } = \int \frac{e^{\lambda/2} }{R} dR = \int \frac{e^{\lambda/2}(Z)}{R(Z)}  R'(Z) dZ,
	\label{tolman}
\end{equation}
and integrating, one obtains:
\begin{equation} 
	\log r = 	\log \left(\sqrt{(M+Z)^2-\frac{M^2}{\gamma ^2}}+M+Z\right) - \log C.
\end{equation}
To make it compatible with (\ref{RZ_})-(\ref{phi_}) the constant of integration should be taken $C=2$. This unambiguously leads to the exact result:
\begin{equation}
	Z = -M +r + \frac{M^2}{4 r \gamma^2 }, 
	\label{Zz}
\end{equation}
from which, in antiscalar limit  $\gamma \to \infty$, we obtain the relation
\begin{equation}
	r = M + Z,
	\label{rMZ}
\end{equation}
which reduces (\ref{RZ_}) to the known transformation law (\ref{PapaTransform}),
and so, together with (\ref{ENU_}) and (\ref{phi_}), implies the transfer from
Fisher's family of solutions again to exponential metric, just as in the JNW and XZ cases.

It is remarkable that equating the scalar charge to mass (in antiscalar regime) in expressions for different coordinate systems leads to the same metric in isotropic coordinates, which, apropos, conforms to observational data not worse than the vacuum Schwarzschild solution \cite{mm18}.

Similar to the Schwarzschild solution having a multitude of extensions (e.g.,
Reisner-Nordstr{\"o}m, Kottler, Kerr, Kerr-Newman, Kerr-AdS, etc.) one can
expect corresponding (horizon-free) alternatives extending from the exponential Papapetrou metric, with the most important and non-trivial extension related to the rotational systems.

\section{Conclusion}
We have expressed the three analytically different presentations -- the JNW, XZ and Fisher families of solutions -- through the single master $\gamma$-factor~(\ref{gamma}) containing the scalar charge in two distinct regimes (scalar and antiscalar), and proved that in the special limit $\gamma \to \infty$, corresponding to the equality of scalar charge and mass in antiscalar regime, all three mentioned families reduce to the same exponential Papapetrou metric. Thereby we have specified the value for the scalar charge which now might be verified, e.g., on the basis of the lensing effect as proposed, e.g., in \cite{Virbhadra1998}, or in some different way (see, e.g., \cite{mm18} and \cite{sau2020}). This being the case, a key consequence of such approach is that all masses might be considered as natural scalar field sources, and thus one can deduce the fundamental character of such self--gravitating background field in nature.

\begin{acknowledgements}
The research was performed within the program No. BR05236322 by the Ministry of Education and Science of the Republic of Kazakhstan.
\end{acknowledgements}

% \section*{Conflict of inte%rest}

\bibliographystyle{spphys}       % APS-like style for physics
\bibliography{references}   % name your BibTeX data base


\end{document}